\documentclass[twocolumn]{layout/tudelft-aiaa}

\usepackage[T1]{fontenc}
\usepackage[utf8]{inputenc}

\usepackage{graphicx} 
\usepackage{amsmath} 
\usepackage{siunitx} 
\usepackage{tabularx} 
\usepackage{subcaption} 

\title{Multi-modal data generation with a deep metric variational autoencoder}

\author[1]{Josefine Vilsbøll Sundgaard}
\author[1]{Morten Rieger Hannemose}
\author[2]{Søren Laugesen}
\author[3]{Peter Bray}
\author[2]{James Harte}
\author[4]{Yosuke Kamide}
\author[5]{Chiemi Tanaka}
\author[1]{Rasmus R. Paulsen}
\author[1]{Anders Nymark Christensen}

\affil[1]{Department of Applied Mathematics and Computer Science, Technical University of Denmark, Denmark}
\affil[2]{Interacoustics Research Unit, c/o Technical University of Denmark, Denmark}
\affil[3]{Interacoustics A/S, Middelfart, Denmark}
\affil[4]{Kamide ENT clinic, Shizuoka, Japan}
\affil[5]{Diatec Japan, Kanagawa, Japan}

\begin{document}

\AlwaysPagewidth{

\maketitle


\begin{abstract}
\noindent We present a deep metric variational autoencoder for multi-modal data generation. The variational autoencoder employs triplet loss in the latent space, which allows for conditional data generation by sampling in the latent space within each class cluster. The approach is evaluated on a multi-modal dataset consisting of otoscopy images of the tympanic membrane with corresponding wideband tympanometry measurements. The modalities in this dataset are correlated, as they represent different aspects of the state of the middle ear, but they do not present a direct pixel-to-pixel correlation. The approach shows promising results for the conditional generation of pairs of images and tympanograms, and will allow for efficient data augmentation of data from multi-modal sources.
\end{abstract}
}

\section{Introduction}
\label{sec:intro}

Deep generative models can generate new data within the distribution of the training dataset, and can be used for advanced data augmentation in cases where data are costly to annotate or difficult to acquire \cite{shin2018medical}. A widely used model is the variational autoencoder (VAE). The VAE is a probabilistic model, consisting of an encoder that learns an approximation of the posterior distribution of the data, and a decoder that learns to reconstruct the original input from a latent representation. An advantage of VAEs over generative adversarial networks (GANs) is that the VAE learns a smooth latent representation of the input data. The latent space can therefore be used for sampling new latent representations and thus be used to generate new examples from the distribution of the training dataset.

Conditional data generation, e.g.,\ the conditional VAE \cite{Kingma2014a}, allows us to specify which class in the dataset to generate data from. Here, both the latent representations and the input data are conditioned by, e.g.,\ class label. Instead of conditioning the model for class specific data generation, Karaletsos et al.~\cite{Karaletsos2016} proposed the triplet-loss based VAE for generation of interpretable latent representations that separate the classes in the latent space with deep metric learning. Karaletsos et al.~\cite{Karaletsos2016} put their main focus on learning the latent representations, whereas we are interested in using the triplet-loss based VAE for data generation.

\begin{figure*}[t]
\centering
\includegraphics[width=1\textwidth]{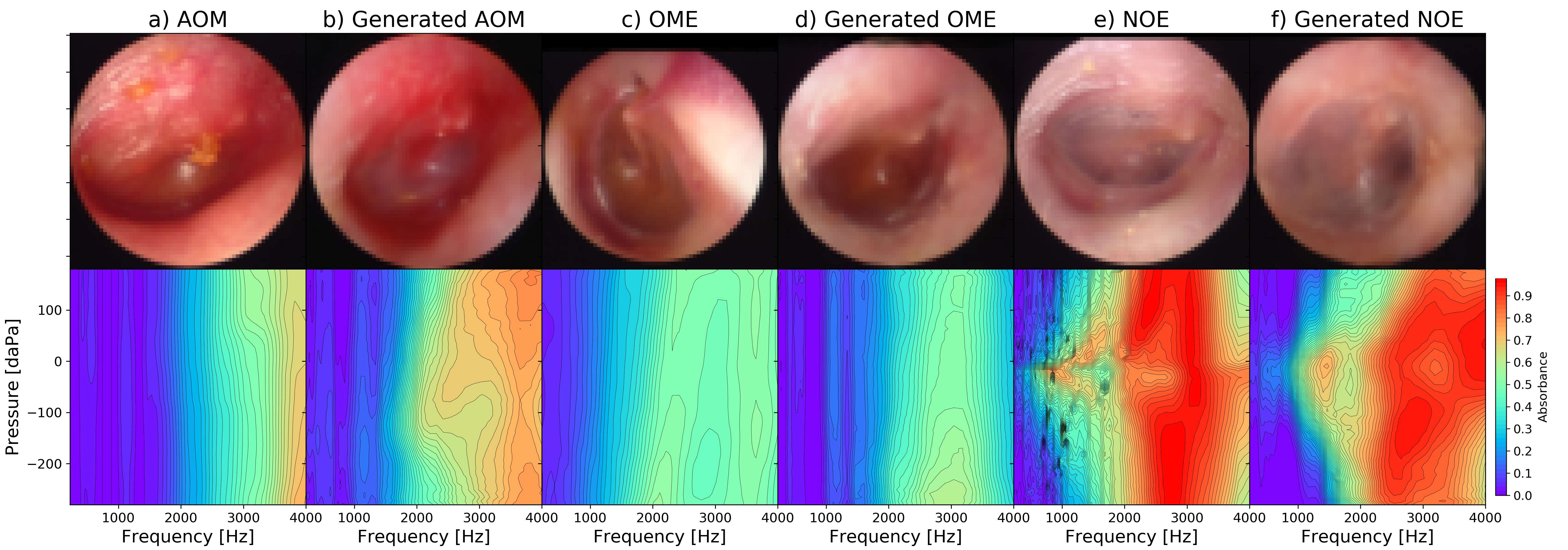} 
\caption{Examples from the dataset and generated examples: otoscopy images (top) and WBT measurements (bottom). Acute otitis media (left two images), otitis media with effusion (middle two images), no effusion (right two images).}
\label{fig:data_original}
\end{figure*}

We propose a generative approach using a triplet-loss based VAE, and we expand the network architecture and training process to allow for multi-modal data generation. The multi-modal dataset consists of pairs of otoscopy images of the tympanic membrane and wideband tympanometry (WBT) measurements, examples of which are presented in Figure \ref{fig:data_original}. The two types of data are very different, as the first is an image from a camera, and the other is the results of an acoustic measurement. Furthermore, they reflect different aspects of the state of the middle ear. The otoscopy image shows the visual impression of the tympanic membrane, which can show signs of e.g.\ infection or effusion, while the WBT measurement provides quantitative indications about the presence of fluid in the middle ear, the mobility of the tympanic-ossicular system, and the volume of the external auditory canal. The two types of data are therefore correlated but do not have a direct pixel-to-pixel relation, and they reflect two different aspects of the state of the middle ear.

Otitis media can be separated into two main diagnostic groups: acute otitis media (AOM) and otitis media with effusion (OME). Figure~\ref{fig:data_original} shows the difference between these two groups, where AOM is an acute infection with redness and a bulging eardrum, and OME is a build-up of fluid in the middle ear. An example of a normal eardrum with no effusion (NOE) is also shown. The WBT measurements in Figure~\ref{fig:data_original} show how the absorbance across the pressure axis does not change in AOM or OME measurements, whereas the NOE measurements typically show a general increase in absorbance around 0 daPa, compared to negative or positive relative pressures. Furthermore, the general absorbance level at lower frequencies is lower for AOM and OME, than for NOE measurements. These two types of data can both be used for the diagnosis of otitis media. Several studies have developed different approaches for otitis media classification based on either otoscopy images~\cite{Senaras2018, Sundgaard2021, Myburgh2018} or WBT measurements~\cite{Grais2021, Terzi2015}. A combined deep learning classification approach based on standard single-frequency tympanograms and otoscopy images was proposed by Binol et al.~\cite{Binol2020}. 

The aim of this paper is to generate new pairs of otoscopy images and WBTs from each of the three diagnostic groups: AOM, OME, and NOE, and for this task, we propose the multi-modal triplet VAE. The generated otoscopy image and WBT pairs can be used as advanced data augmentation for a multi-modal classification pipeline. Our multi-modal generative model can also be used in other domains such as pairs of cardiac images and electrocardiograms, or brain scans and electroencephalograms. These modalities have a correlation, while reflecting different aspects - visual and functional - of the condition of the examined organ. This work can also be used for the training of doctors and models while preserving patient privacy. Generated data ensures anonymity and allows for data to be shared without regulations such as EU's GDPR, and some studies have already shown the usability of variational autoencoders in this field \cite{li2019evaluating, shin2018medical}.

\section{Methods}
\begin{figure}[t]
\centering
\includegraphics[width=0.48\textwidth]{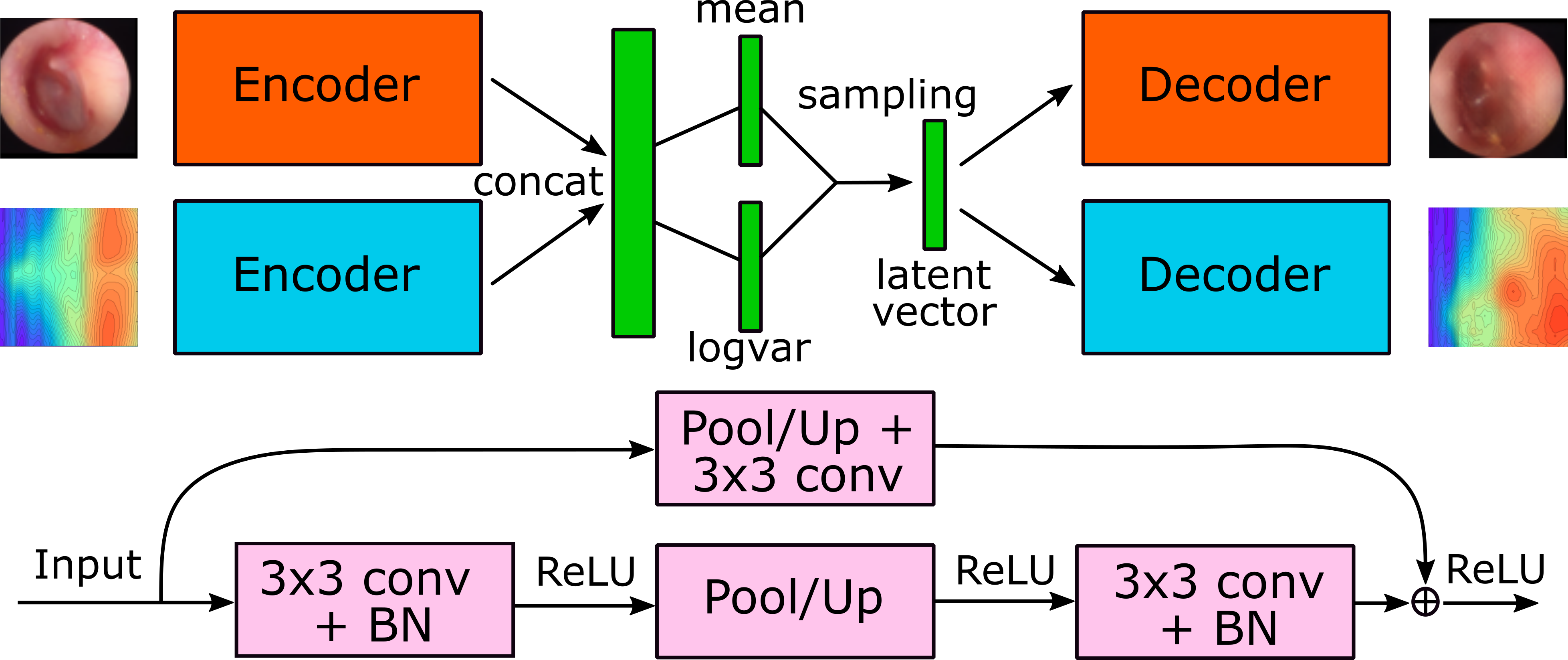} 
\caption{Structure of the multi-modal triplet VAE. Top figure shows the overall structure with two encoders, concatenation of the outputs, sampling, and two decoders. Bottom figure shows the residual blocks used in both encoders and decoders. BN refers to batch normalization.}
\label{fig:tripletVAE}
\end{figure}
The multi-modal triplet VAE consists of two encoders and two decoders - one for each modality, and the structure is shown in Figure~\ref{fig:tripletVAE} together with the structure of the upsampling and downsampling blocks used to construct the encoders and decoders. The encoders consist of five residual downsampling blocks using 2D average pooling, and take the $64\times64\times3$ otoscopy images and the $64\times64\times1$ WBT measurements as input. They start with 64 features in the first block, and double the number of features in each consecutive block. The output feature maps from each encoder ($2\times2\times512$) are concatenated, and two $2\times2$ convolutional layers are used to obtain the mean and variance in the 128-dimensional latent space. Using the reparameterization trick~\cite{Kingma2014a}, a latent vector is sampled, which is passed to both decoders. The decoders consist of six residual upsampling blocks using nearest neighbour upsampling, and the number of features is halved for each block starting at 512. The final layer is a single $3\times3$ convolutional layer going from 32 feature maps to the desired number of channels of the output - one channel for WBTs and three for the otoscopy images. Because the encoder outputs are concatenated, we achieve a common latent space for both modalities, which allows for sampling in the latent space to generate new pairs from each class. The decoders will thus receive information from both image and WBT for the reconstruction of each modality.

The training loss function consists of several parts. The difference between reconstructed WBT and input WBT is penalized using binary cross entropy (BCE) loss. The reconstruction of the image is evaluated using structured similarity index (SSIM) loss~\cite{wang2004image}, which is a local measurement comparing the reconstruction and original image based on luminance, contrast, and structural information. In the latent space, both Kullback–Leibler (KL) divergence and triplet loss \cite{Schroff2015} are computed. The KL divergence forces the latent embeddings close to a standard normal distribution, while the triplet loss forces examples from the same class to cluster together and pushes examples from different classes further apart~\cite{Schroff2015}. The loss function terms related to the embedding space are weighted lower than the rest of the terms, and the value 0.1 was experimentally chosen, leading to a loss function defined as:
\begin{equation}
Loss = L_{SSIM} + L_{BCE} + 0.1 \cdot ( L_{KL} + L_{triplet})
\end{equation}

Balanced sampling is performed during training, with a batch size of 60 (20 pairs from each class) to ensure a balanced representation of every class in each training batch and to cope with the class imbalance in the dataset. The triplets are sampled in each batch using semi-hard mining \cite{Schroff2015} based on the encoder-generated mean vector from each input pair. The VAE is trained for 5000 epochs using the Adam optimizer~\cite{Kingma2014} with a learning rate of $0.0004$. Data augmentation is performed using random erasing~\cite{Zhong2020} on both image and WBT measurement, while horizontal flipping and rotation with $\pm 20$ degrees is also performed on the images.

Once the network is trained, the test set is passed through the encoders, obtaining the latent representation of each image and WBT pair in the test set. In order to sample new latent vectors for generation of data pairs in each class, the distribution of each class in the latent space is approximated using kernel density estimation for each class. Kernel density estimation estimates the probability density function in the latent space by placing a Gaussian kernel on each sample. The bandwidth of the kernel is fine-tuned using five-fold cross validation. The kernel density estimation is performed only on the test set. When the distribution of each class is estimated, new samples can be generated. The sampled latent vectors are then run through both decoders, to generate new pairs of images and WBTs.

\subsection{Data}
The dataset consists of 1420 pairs of images and WBT measurements collected at Kamide ENT clinic, Shizouka, Japan, from patients aged between 2 months and 12 years. Each pair was assigned one of the three classes: NOE (537 pairs), OME (419 pairs), and AOM (211 pairs) by an experienced ENT specialist based on signs, symptoms, patient history, otoscopy examination, and WBT measurements. The data was collected and handled under the ethical approval from the Non-Profit Organization MINS Institutional Review Board (reference number 190221), with either opt-out consent, or informed consent from all participants or their parent or guardian.

An otoscopy image is captured using an endoscope (dedicated video otoscope) inserted into the ear canal, allowing a visual inspection of the tympanic membrane. The original image size was $640\times 480$ pixels, which was cropped to a square to limit the amount of black background and then downsampled to $64\times64$ to fit the proposed architecture. A WBT measurement is performed by inserting and hermetically sealing an acoustic probe with an appropriately sized silicone ear tip into the patient's ear canal. The probe repeatedly presents a transient stimulus with a frequency range encompassing 226 Hz to 8 kHz while modifying the pressure in the external acoustic canal relative to the ambient pressure from 200 to -300 daPa~\cite{Hein2017}. The measurements were performed using the Titan system (Interacoustics, Denmark). From the WBT measurement, it is possible to derive conclusions about both tympanic membrane mobility and middle ear condition, and thus additional diagnostic power can be gained over visual inspection alone. WBT measurements were bilinearly resampled to a common grid from 180 daPa to -280 daPa in 64 steps on a linear scale for the pressure axis, and from 226 Hz to 4 kHz in 64 steps for the frequency axis. Examples of both images WBT measurements are shown in Figure~\ref{fig:data_original}.

The dataset is split into a train (80\%) and test (20\%) set. It was ensured that data from one patient was only used for either training or testing, to prevent data leakage.

\section{Results}
\begin{figure}[b!]
\centering
\includegraphics[width=0.45\textwidth]{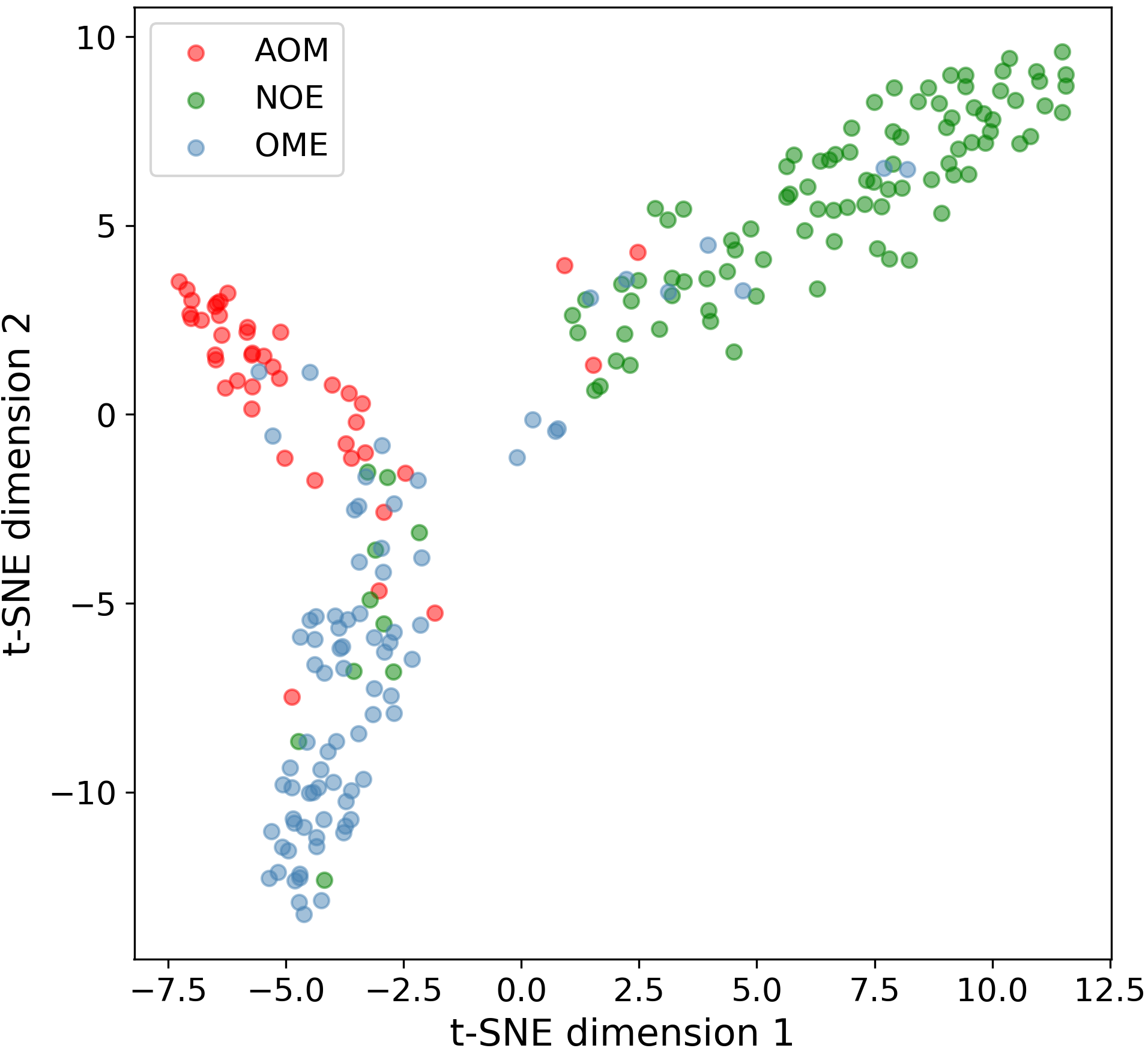} 
\caption{t-SNE visualization of test data latent embeddings.}
\label{fig:embeddings}
\end{figure}

The test embeddings are shown in Figure~\ref{fig:embeddings}. The 128-dimensional latent representation of each image has been reduced to two dimensions using t-SNE dimensionality reduction \cite{VanDerMaaten2008} in order to visualize the latent space. The test embeddings clearly show three clusters, but they do blend in the transition areas between the classes, as the images and WBTs can look quite similar across the diagnostic groups. Some of the overlap could also arise from the drastic dimensionality reduction from 128 to two dimensions. The clusters will likely be more separable in the high-dimensional space.

\begin{figure}[]
\centering
\includegraphics[width=0.48\textwidth]{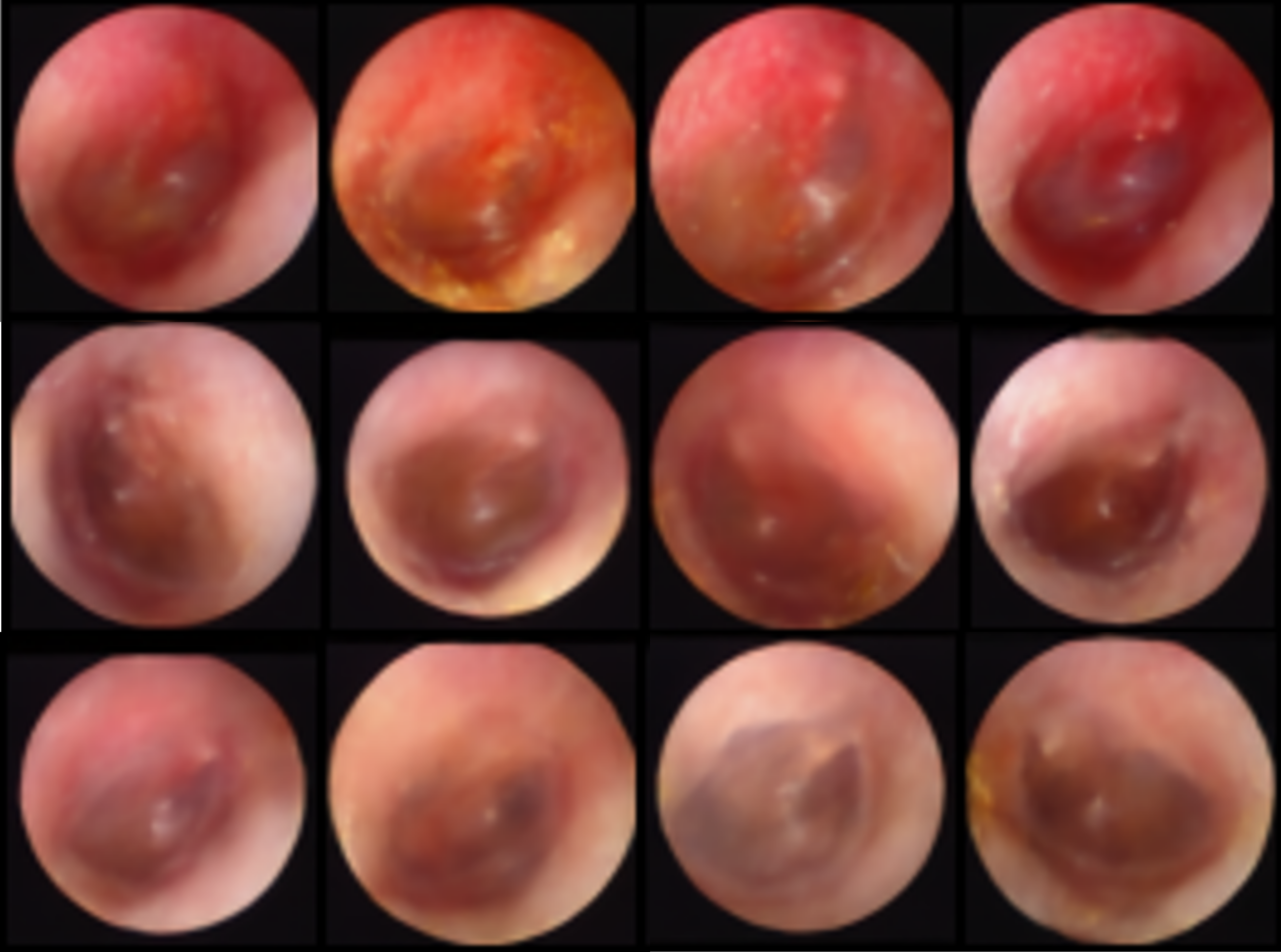} 
\caption{Examples of generated otoscopy images. Top row: AOM, middle row: OME, bottom row: NOE. Best viewed with zoom.}
\label{fig:imggen}
\end{figure}

\begin{figure}[]
\centering
\includegraphics[width=0.48\textwidth]{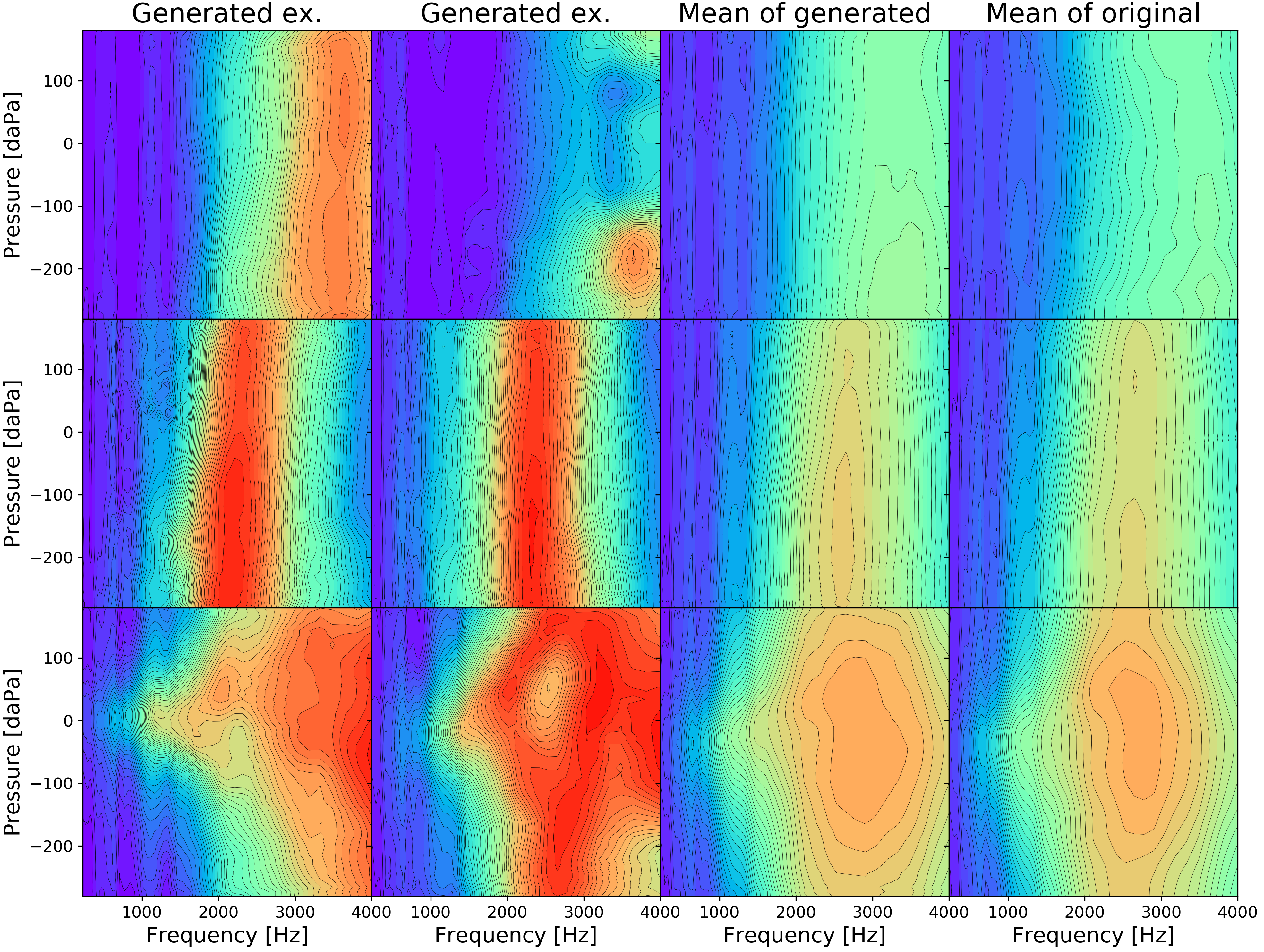} 
\caption{Overview of generated WBT measurements. Top row: AOM, middle row: OME, bottom row: NOE. Best viewed with zoom.}
\label{fig:WBTgen}
\end{figure}

New latent representations are sampled in the full 128-dimensional space within the three class distributions estimated with kernel density estimation, and examples of generated otoscopy images and WBTs are plotted in Figures~\ref{fig:imggen} and \ref{fig:WBTgen}.

Figure~\ref{fig:imggen} shows examples of generated images in the three diagnostic groups. The images look realistic, as they all contain a tympanic membrane, clear diagnostic markers, and the malleus bone is seen in several examples. The top row of AOM images shows signs of redness and bulging eardrum, and the OME cases clearly have effusion behind the eardrum. The NOE cases appear pale and translucent, as expected.

Other examples of generated pairs of otoscopy images and WBTs are shown side by side with original examples from the dataset in Figure~\ref{fig:data_original}. These are not reconstructions, but new generated images. In this figure, it is possible to compare the diagnostic markers of the conditions across modalities, while also comparing the generated examples with original examples. Figure~\ref{fig:data_original}(a-b) show similar signs of AOM redness and infection and reduced absorbance in the WBT, which is relatively flat across the pressure axis. The two OME cases in Figure~\ref{fig:data_original}(c-d) show very similar diagnostic signs on both the original and generated data with yellow effusion behind the tympanic membrane. Likewise, the absorbance is much lower with very little variation across pressures. The NOE cases in Figure~\ref{fig:data_original}(e-f) show normal tympanic membranes and high absorbance in the WBT with a change across pressure.

The generation of WBT measurements is summarized in Figure~\ref{fig:WBTgen}, where generated examples are shown together with the average WBT of the generated samples as well as the original dataset for each of the three diagnostic groups. The average of the generated samples is computed from 500 samples in each diagnostic group. The two average WBT measurements look very similar. This shows that the generated WBT measurements within each diagnostic group follow the same pattern as the mean of the original dataset, thus the distribution of the classes has been captured quite well. The generated examples also indicate great variation within each class. 

\section{Discussion and Conclusion}
The proposed multi-modal triplet-loss based VAE is able to generate highly realistic conditional pairs of otoscopy images and WBT measurements. The generated images examples in Figures~\ref{fig:data_original} and \ref{fig:imggen} show that the proposed triplet-loss based VAE generates images with a large variation in appearance, and with clear diagnostic markers. The generated images does appear a bit blurry, which is a common VAE problem~\cite{Dosovitskiy2016}. The use of SSIM loss \cite{wang2004image} has improved the quality of the generated images drastically, compared to employing BCE loss. Other studies have found ways to improve the quality even further, and have thus synthesized high resolution images using VAEs~\cite{Huang2018, Zhao2017}. However, incorporating this into our approach remains future work. The WBT is a simpler type of data to generate, as it does not contain the same level of detail as an image. BCE loss is therefore sufficient for this modality, and the results in Figures~\ref{fig:data_original} and \ref{fig:WBTgen} show that the generated WBTs corresponds very well to the appearance and structure of the original WBTs.

In this study, we propose a VAE structure for conditional multi-modal data generation, even when no direct pixel-to-pixel correlation is present in the two modalities. This multi-modal VAE structure is very flexible, as the encoder and decoder for each modality are completely de-coupled from the other modality. This allows different architectures to be used for each modality depending on the specific needs of the modalities. The employed network architecture for the otoscopy images could be changed to allow for generation of larger and more high-quality images. Likewise, the architecture could be altered to fit temporal data, such as electrocardiograms or electroencephalograms, if this method was to be employed in other domains.

Furthermore, the results show how conditional data generation can be accomplished when employing triplet loss in the latent space of the VAE. This way, conditioning the input or latent space is not needed, as one can simply sample within a certain class cluster. 

\section{Acknowledgements}
This study was financially supported by William Demant Foundation.

\bibliography{LiteratureReview}

\end{document}